\documentclass{emulateapj}
\def\stacksymbols #1#2#3#4{\def\theguybelow{#2}
        \def\verticalposition{\lower#3pt}
        \def\spacingwithinsymbol{\baselineskip0pt\lineskip#4pt}
        \mathrel{\mathpalette\intermediary#1}}
\def\intermediary #1#2{\verticalposition\vbox{\spacingwithinsymbol
        \everycr={}\tabskip0pt
        \halign{$\mathsurround0pt#1\hfil##\hfil$\crcr#2\crcr
                \theguybelow\crcr}}}
\def\lta{\stacksymbols{<}{\sim}{2.5}{.2}}
\def\gta{\stacksymbols{>}{\sim}{3}{.5}}

\begin{document}

\title{STOPPING COOLING FLOWS WITH COSMIC RAY FEEDBACK}




\author{
William G. Mathews\footnotemark[1]
}

\footnotetext[1]{UCO/Lick Observatory,
Dept. of Astronomy and Astrophysics,
University of California, Santa Cruz, CA 95064}



\begin{abstract}
Multi-Gyr two-dimensional
calculations describe the gasdynamical evolution of 
hot gas in the Virgo cluster resulting from intermittent 
cavities formed with cosmic rays. 
Without cosmic rays, the gas evolves into a 
cooling flow, depositing about 85 solar masses
per year of cold gas in the
cluster core -- such uninhibited cooling conflicts with
X-ray spectra and many other observations.
When cosmic rays 
are produced or deposited 
10 kpc from the cluster center in bursts 
of about $10^{59}$ ergs lasting 
20 Myrs and spaced at intervals of 200 Myrs, 
the central cooling rate is greatly reduced to
${\dot M} \approx$ 0.1 - 1 solar masses per
year, consistent with observations. 
After cosmic rays diffuse through the cavity walls, 
the ambient gas density is reduced and is buoyantly 
transported 30-70 kpc out into the cluster.
Cosmic rays do not directly heat the gas 
and the modest shock heating around young cavities 
is offset by global cooling as the cluster gas expands.
After several Gyrs the hot gas density and temperature profiles
remain similar to those observed,
provided the time-averaged cosmic ray luminosity is about
$L_{cr} = 2.7 \times 10^{43}$ erg s$^{-1}$, 
approximately equal to the bolometric cooling rate $L_X$ within
only $\sim56$kpc.
If an appreciable fraction of the relativistic cosmic rays are
protons, gamma rays produced by pion decay
following inelastic p-p collisions
may be detected with the Fermi Gamma Ray Telescope.
\end{abstract}


\section{Introduction}
X-ray spectra of hot virialized gas in galaxy groups and clusters 
have firmly established that the rate that gas cools to low 
temperatures is much lower than that expected from 
uninhibited radiative cooling in subsonically inflowing 
cooling flows  
(e.g. Peterson et al. 2001). 
As a result of this realization, 
many theoretical and computational studies have 
sought to prevent cooling inflows by introducing a 
gas heating mechanism usually exploiting the 
phenomenal accretion energy released by 
massive black holes in the cores of group and 
cluster-centered elliptical galaxies 
(e.g. McNamara \& Nulsen 2007). 
In spite of this almost unlimited source of accretion 
energy, 
no specific mechanism or mode of 
gas heating has been accepted 
that is successful for long cosmic times and for 
cooling flows of all scales: galactic, group and cluster.

In their recent review of the cooling flow puzzle 
McNamara \& Nulsen (2007) (MN07)
emphasize two types of AGN-related heating mechanisms: 
$PdV$ work done when cavities in the hot 
gas are inflated by cosmic rays (CRs) and 
energy-dissipating shock (or other) waves that deliver energy 
from the central energy source throughout the cluster gas. 
CRs, often observed inside cluster cavities from  
radio synchrotron emission, 
are thought to be deposited or produced by non-thermal jets 
that project out from the central AGN.
However, our recent study of the energetics of cluster 
cavities inflated by CRs 
(Mathews \& Brighenti 2008b = MB08)
shows that the (shock) heating produced as young cavities 
expand due to locally enhanced CRs 
is more than compensated by a global cooling 
as the entire cluster gas expands more or less permanently 
to accommodate the partial pressure of the CRs 
as they diffuse into the cluster gas. 
Most of the $PdV$ work expended as cavities form is 
stored in potential energy of displaced 
gas that moves out 
in the cluster potential. 
The dissipative heating beneath buoyant
cavities discussed in MN07 appears to be absent in the 
cavities computed in MB08, and this is consistent with our earlier 
estimate (Mathews et al. 2006). 
The total (bulk and turbulent) kinetic energy 
is always very small and cannot be an important source 
of dissipative heating (MB08).
%
%
%

The second currently favored heating mechanism, 
large scale wave dissipation, is also problematic. 
Since the hot gas density profiles in groups and clusters is
generally flatter than $\rho \propto r^{-2}$,
successive shocks
preferentially heat gas near the cluster center 
until it exceeds observed temperatures 
(Mathews, Faltenbacher \& Brighenti 2006).
Over-heating the central gas 
is a persistent difficulty with 
previous theoretical attempts to
resolve the cooling flow problem by 
using central AGN heating scenarios 
(Brighenti \& Mathews 2002b, 2003) 
since the cluster gas closest to the source of AGN energy
typically has the lowest temperatures
observed in the hot cluster gas.
Finely-tuned heating scenarios in which the gas inflow is
assumed to be essentially stopped are inconsistent with 
secular changes in the gas that evolve without limit. 
For example metal enrichment by Type Ia supernovae in
the central galaxy would eventually  
increase the local hot gas metal abundances 
far above those observed.

As an alternative solution to the cooling flow problem, 
we proposed that 
gas receiving AGN energy is buoyantly 
transported outward. 
In these mass-circulating flows 
gas flows in both radial directions simultaneously.
The combined flows  
maintain a cooling flow appearance 
similar to observed X-ray images  
(Mathews et al. 2003, 2004;
Brighenti \& Mathews 2006).

In this Letter we describe the evolution of 
virialized hot gas in the Virgo cluster in which 
successive cavities are formed with CRs 
that buoyantly transfer gas from near the cluster core 
to large radii where 
its radiative cooling time is very much longer.
The average net gas mass that flows past every 
cluster radius is very small although 
significant inflows and outflows occur locally 
in the gas. 
In buoyant outflows generated by CRs, 
the cluster gas need not be directly heated 
by the CRs, 
although some modest 
heating results from the dissipation of weak shocks 
that propagate away from newly formed cavities 
(Mathews \& Brighenti 2007; MB08). 
Unlike most AGN heating scenarios, 
buoyant mass-circulating flows generated by CRs 
can preserve the observed cluster 
temperature and density profiles for at least several Gyrs. 
This resolution of the cooling flow problem, 
using mass transfer by CR buoyancy rather 
than heating, 
is attractive because almost 
all group and cluster-centered galaxies are observed to 
produce CRs in jets, radio lobes, or in 
central non-thermal radio sources. 

\section{Equations and Procedure}

The procedure and assumptions employed here 
are nearly identical to those described in detail by 
MB08 to which the reader is referred. 
The flow equations are 
\begin{equation}
{ \partial \rho \over \partial t}
+ {\bf \nabla}\cdot\rho{\bf u} = 0
\end{equation}
\begin{equation}
\rho \left( { \partial {\bf u} \over \partial t}
+ ({\bf u \cdot \nabla}){\bf u}\right) =
- {\bf \nabla}(P + P_c) - \rho {\bf g}
\end{equation}
\begin{equation}
{\partial e \over \partial t}
+ {\bf \nabla \cdot u}e = - P({\bf \nabla\cdot u})
- (\rho/m_p)^2 \Lambda(T,z)                                                           
\end{equation}
\begin{equation}
{\partial e_c \over \partial t}
+ {\bf \nabla \cdot u}e_c = - P_c({\bf \nabla\cdot u})
+ {\bf \nabla\cdot}(\kappa{\bf \nabla}e_c)
+ {\dot S}_c
\end{equation}
where artificial viscosity terms are suppressed. 
Gas and CR pressures are related to their 
corresponding energy densities by 
$P = (\gamma -1)e$ and $P_c = (\gamma_c - 1)e_c$
respectively where for simplicity 
we assume $\gamma  = 5/3$ and $\gamma_c = 4/3$ 
for fully relativistic CRs.
The energy spectrum of the CR 
particles $N(E)$ is not considered in detail 
so the CRs are characterized only by their
integrated energy density $e_c \propto \int EN(E)dE$.
Furthermore, the CR particles are unspecified 
and can consist of electrons or protons in any combination. 

Unlike the discussion in MB08, 
we now include radiative cooling. 
We adopt the radiative cooling 
coefficient $n_{ions}n_e\Lambda_{sd}(T,z)$ 
erg cm$^{-3}$ s$^{-1}$ from
Sutherland and Dopita (1993) which, 
as expressed in equation 3, 
$(\rho/m_p)^2 \Lambda$ erg cm$^{-3}$ s$^{-1}$, 
requires that 
$\Lambda = 1.1[(4 - 3\mu)(2 + \mu)/25\mu^2]\Lambda_{sd}$
where $\mu = 0.61$ is the molecular weight 
and the coefficient is $n_{ions}/n_p = 1.1$.
We do not compute the metallicity 
in the gas throughout the flow but simply 
adopt a uniform metallicity near solar $z = 0.75z_{\odot}$ 
since most of the gas that cools has been metal-enriched in 
the cluster core.
The computed flows are not sensitive to 
the gas metallicity. 
Computational timesteps are required to be less than 
the constant-pressure radiative cooling time in all zones. 
Since some cooling occurs in off-center zones near the $z$-axis, 
it is useful to remove cooled gas  
from the computational grid.
If a grid zone cools below $T_{min} = 5 \times 10^5$ K,
gas is removed at constant pressure 
by reducing the zone density until 
the temperature resets to $T_{max} = 5 \times 10^6$ K. 
Gas removal at constant pressure from a zone preserves the 
pressure gradients relative to nearby zones so the gas flow to or 
from neighboring zones is unaffected. 
Moreover, the total thermal energy in the zone is also preserved so 
thoroughly cooled gas removed in this way does not 
alter the total thermal energy of uncooled gas.
We assume that CRs are not removed with the cooled gas. 

Gas is accelerated by pressure gradients in 
both the thermal gas and CRs.
The two fluids are assumed to be coupled by the presence of 
small magnetic fields in the cluster gas. 
This coupling can be efficient even 
when the energy density in the magnetic field is much less 
than that in the thermal gas, which is consistent with 
the small fields observed in clusters, $1 - 10\mu$G 
(Govoni \& Feretti 2004). 
Consequently, magnetic pressure and stresses 
are not explicitly considered. 

Equation 4 above describes the advection of 
CRs by locally flowing gas and the diffusion 
of CRs through this gas. 
Little is known about the diffusion coefficient 
$\kappa$, but it is likely that it varies inversely 
with the gas density since the magnetic field is 
probably larger in denser gas. 
As in MB08 
we consider a range of CR diffusion coefficients 
that depend on a a single density parameter $n_{e0}$:
\begin{displaymath}
\kappa = \left\{
\begin{array}
{r@{\quad:\quad}l}
10^{30} ~{\rm cm}^2 {\rm s}^{-1} & n_e \le n_{e0} ~{\rm cm}^{-3} \\
10^{30}(n_{e0}/n_e) ~{\rm cm}^2 {\rm s}^{-1} & n_e > n_{e0} ~{\rm
  cm}^{-3}
\end{array} \right.
\end{displaymath}
Here we consider 
$n_{e0} = 6 \times 10^{-3}$ and $6 \times 10^{-6}$ cm$^{-3}$. 
The cluster gas dynamics 
are not very sensitive to $n_{e0}$ or to other 
$\kappa(n_e)$ variations that have been explored
(MB08). 

The flow equations are solved in 2D cylindrical coordinates 
$(R,z)$ using a ZEUS-like code 
(Stone \& Norman 1992) but with the additional 
CR equation. 
As in MB08 the computational grid 
consists of 100 equally-spaced zones out to 50 kpc 
plus an additional 100 zones in both coordinates that 
increase logarithmically out to $\sim 1$ Mpc, 
filling a large hemisphere. 
Also following MB08, 
a spherical gravitational potential for 
the Virgo cluster is found 
by assuming that the gas is in hydrostatic equilibrium 
with the temperature and modified density profiles $T_G(r)$ 
and $\rho_{mG}(r)$
observed by Ghizzardi et al. (2004). 
The density profile was found by integrating the 
equation of hydrostatic equilibrium using $T_G(r)$ 
and the total gravitational acceleration from 
a $3.2 \times 10^9$ $M_{\odot}$ black hole, 
the M87 stellar mass $\rho_*(r)$ from 
Brighenti \& Mathews (2002a), and an NFW profile 
with $M_{vir} = 1.30 \times 10^{14}$ $M_{\odot}$ 
and concentration 8.60.
The resulting modified observed density profile 
$\rho_{mG}(r)$ agrees very well 
with that of Ghizzardi et al. $\rho_G(r)$ for $r \gta 1$ kpc 
but avoids an unphysical maximum in $g(r)$
near $\sim 4$ kpc if $g(r)$ is found 
from $T_G(r)$ and $\rho_G(r)$.
All flow calculations begin with  $T_G(r)$
and $\rho_{mG}(r)$.

CRs are introduced 
at intervals of 
$t_{cyc} = 2 \times 10^8$ yrs 
in bursts lasting 
$t_{cav} = 2 \times 10^7$ yrs.
These timescales are roughly consistent with the lifetimes 
of visible X-ray cavities and radio synchrotron electrons. 
If the CR luminosity during injection times is 
${\dot E}_c$, the time-averaged CR luminosity is 
\begin{equation}
\langle L_{cr} \rangle = 
2(t_{cav} / t_{cycle}) {\dot E}_c, 
\end{equation}
assuming that 
CRs are injected into both cluster hemispheres 
by symmetric double jets.
For comparison with the cavity calculations 
in MB08, 
all cavities are assumed to form in a source region  
10 kpc along the $z-$axis 
described by a normalized Gaussian profile. 
The CR source term in equation 4 is therefore 
\begin{equation}
{\dot S}_c = {\dot E}_c \tau(t)
{e^{-(({\bf r - r}_{cav})/r_s)^2} \over \pi^{3/2} {r_s}^3}
~~~{\rm erg}~{\rm  cm}^{-3}{\rm s}^{-1}
\end{equation}
where ${\bf r}$ is the distance from the origin, $r_s = 2$ kpc 
and $\tau(t)$ is unity during injection times and 
zero otherwise. 
All calculations described here continue for 3 Gyrs.
For each assumed $\kappa(n_{e0})$ we vary ${\dot E}_c$, 
seeking that value for which the gas temperature 
and density profiles continue to agree best with 
profiles observed in Virgo.
Except for radiative cooling and a variable ${\dot E}_c$,
the parameters chosen here are
identical to those discussed for single cavities by MB08.

\section{Results}

The left column of Figure 1 illustrates the evolution of 
Virgo cluster gas 
as a pure radiatively cooling flow without CRs.
The long-dashed lines 
show the radial gas density, temperature 
and pressure profiles after evolving for 3 Gyrs 
from the observed profiles. 
The evolved profiles are determined by ordering 
$\rho$, $T$ and $P$ in each 
computational zone by increasing distance from the 
origin $r = (R^2 + z^2)^{1/2}$, 
then averaging in appropriate bins $\Delta r$. 
After 3 Gyrs, $\rho(r)$ and $T(r)$ 
have evolved away from the initial (observed) profiles 
(solid lines). 
The ratio of gas entropy to its initial value 
(subscript 0), 
$\log(S/S_0) = \log(T/T_0) - (2/3)\log(\rho/\rho_0)$, 
decreases monotonically toward the cluster center 
where the gas cools in the central 1.5 kpc.
The discrepancy in integrated mass of (uncooled) gas 
between the initial and evolved
density profiles in the uppermost panel, 
$\Delta M = M(r,3{\rm~Gyr}) - M(r,0)$, 
peaks at $\sim 15$ kpc\footnotemark[3].
\footnotetext[3]{This suggests the approximate site to 
initiate buoyant outflow.}
Between 2 and 3 Gyrs the quasi-steady cooling rate in the core 
is $\langle {\dot M} \rangle = 86$ $M_{\odot}$ yr$^{-1}$ 
for the entire Virgo cluster (both hemispheres)\footnotemark[4].
\footnotetext[4]{The mass cooling rate, found from the total 
mass of cooled gas removed during this time interval, 
is insensitive to $T_{min}$ and $T_{max}$. For example,
$\langle {\dot M} \rangle = 85$ $M_{\odot}$ yr$^{-1}$ 
with $T_{min} = 3 \times 10^5$ K and $T_{max} = 10^6$ K.}

Panels in the central column of Figure 1 show the 
corresponding evolution with CRs having a low 
CR diffusion coefficient 
$\kappa$ with $n_{e0} = 6 \times 10^{-6}$ cm$^{-3}$.
In this solution after 3 Gyrs the gas density and temperature 
remain close to the initial, 
observed profiles.
The entropy in 
$25 \lta r \lta 100$ kpc is lowered by the accumulated 
buoyant arrival of gas from the cluster core 
(cf. MB08) that lost entropy by radiation.
The modest entropy maximum near 10 kpc is probably due to
heating by successive weak cavity shocks
at this radius. 
The total mass of hot gas in each spherical annulus 
changes very little during the evolution, 
as expected 
from the small central mass cooling
rate between 2 and 3 Gyrs,
$\langle {\dot M} \rangle = 0.13$ $M_{\odot}$ yr$^{-1}$
(for both hemispheres).
The cooling inflow has been almost completely balanced 
by the outward circulation of gas from the cluster core to 
large radii driven by CR buoyancy.
The bottom panel shows that 
the CR pressure $P_c$ 
after 3 Gyrs is 
about an order of magnitude less than the gas pressure 
within 100 kpc, 
but $P_c/P$ decreases sharply beyond 100 kpc.
In this solution the total (spherical)
time-averaged CR luminosity 
is $\langle L_{cr} \rangle = 2.70 \times 10^{43}$ erg s$^{-1}$. 
The total CR energy delivered to each 
cavity is about $E_{cav} = 8.2 \times 10^{58}$ ergs.

The right column of panels 
in Figure 1 shows the identical 
evolution but with a much larger CR diffusion coefficient 
($n_{e0} = 6 \times 10^{-3}$ cm$^{-3}$).
As before, the gas density and temperature profiles 
remain essentially unchanged after 3 Gyrs 
with a total time-averaged CR luminosity 
$\langle L_{cr} \rangle = 2.60 \times 10^{43}$ erg s$^{-1}$.
The total gas cooling rate between 2 and 3 Gyrs,
$\langle {\dot M} \rangle = 1.15$ $M_{\odot}$ yr$^{-1}$, 
is only $\sim 0.01$ of the cooling flow rate
without CRs. 

\section{Discussion and Conclusions}

CR buoyancy efficiently 
arrests cooling inflows  
without requiring {\it ad hoc} heating 
or infusions of hot, non-relativistic gas in the cavities.
In the two solutions with CRs 
gas cools at the center and along the 
$z$-axis out to about 3 - 5 kpc where the free fall 
time to the center is $\lta 10^7$ yrs.
If all this cooled gas were accreted by the central black hole, 
the efficiency 
required for accretion energy to produce enough CRs 
to maintain the observed gas profiles is 
$\eta_{cr} = \langle L_{cr} \rangle/\langle {\dot M} \rangle c^2$.
We find $\eta_{cr} = 3.6 \times 10^{-3}$ 
and $3.9 \times 10^{-4}$ 
respectively for the low and high $\kappa$ solutions. 
But $\eta_{cr}$ is probably underestimated 
since some of the cooled, dusty gas in group-centered 
elliptical galaxies is heated and buoyantly 
relocated far out into the hot gas 
(Temi, Brighenti, \& Mathews 2007).
The total CR luminosity 
required to (nearly) stop the Virgo cooling flow, 
$\langle L_{cr} \rangle = 2.60 \times 10^{43}$ erg s$^{-1}$, 
is equal to the (computed) bolometric X-ray luminosity 
within only $\sim56$kpc. 

The Fermi Gamma Ray Telescope may confirm that cooling inflows 
in Virgo and other nearby clusters are stopped 
with CR buoyancy. 
Gamma rays are created by neutral pion decays 
following inelastic p-p interactions between 
CRs and the thermal plasma. 
The integrated gamma ray flux (cm$^{-2}$ s$^{-1}$) above 
energy $E_{\gamma,min}$ is 
\begin{equation}
F_{\gamma}(E_{\gamma} > E_{\gamma,min}) = 
{1 \over 4 \pi d^2} \int_{Virgo} dV 
\int_{E_{\gamma,min}}^{\infty} dE_{\gamma} q_{\gamma} (E_{\gamma},r)
\end{equation}
where $d = 16$ Mpc is the distance to Virgo and 
the specific photon emissivity, 
$q_{\gamma}(E_{\gamma},r)$ cm$^{-3}$ s$^{-1}$ GeV$^{-1}$, 
is described by 
Pfrommer \& Ensslin (2004) and Ando \& Nagai (2008).
Assume for simplicity that all CRs are relativistic 
protons having a power law momentum distribution 
$f(p_p,r) \propto p_p^{-\alpha}$ with $\alpha = 2.3$. 
Then after 3 Gyrs the integrated gamma ray photon flux  
from Virgo above $E_{\gamma,min} = 2$ Gev is 
$F_{\gamma} = 2 \times 10^{-9}$ or 
$3 \times 10^{-9}$ cm$^{-2}$ s$^{-1}$ 
respectively for the low and high $\kappa$ solutions 
in Figure 1.
These fluxes are near the detection threshold 
of the Fermi telescope,
$\sim 10^{-9}$ cm$^{-2}$ s$^{-1}$ 
estimated by Ando \& Nagai (2008) for extended sources.
If half of the CRs are electrons, 
evidence of the AGN CRs that we require in  
Virgo may still be marginally visible with the Fermi Telescope.
Additional CRs in Virgo may have been produced in 
shocks and turbulence accompanying mergers or in 
the accretion shock when baryons first fell 
into the Virgo potential 
(Loeb \& Waxman 2000), but these are not considered here. 

For both low and high $\kappa$ solutions in Figure 1
the CR partial pressure $P_c/(P+P_c)$ increases 
toward the cluster center.
This may seem surprising because all CRs 
are injected at 10 kpc and diffusion into the dense, central gas 
is slow because $\kappa$ decreases. 
This feature can be understood from the reaction 
of X-ray cavities on the cluster gas.
As a young cavity buoyantly moves away from the cluster center, 
relatively cooler nearby gas concentrates 
underneath the cavity, 
then flows rapidly outward through the cavity 
along the $z$-axis (MB08).
These jets or filaments of denser thermal gas move out 
beyond 10 kpc then later fall back under gravity 
into the cluster core 
(e.g. Mathews \& Brighenti 2008a). 
It is likely that the central concentration of CRs 
occurs because they are advected in with each infalling 
filament, then become trapped by the relatively small 
$\kappa$ in the dense central gas.
CRs are a likely source of non-thermal 
pressure that is occasionally observed 
in the cores of hot galactic and 
cluster gas.

Figure 2 shows that the distribution of CRs 
in in Virgo after 3 Gyrs 
is concentrated in the core and along the $z$-axis. 
In a recent study of warm and cold gaseous filaments in
the Perseus Cluster, Ferland et al. (2008a,b) find that
the observed spectrum can be understood if some of the 
the emission lines are excited by collisions with 
CRs.
The enhanced CR distribution 
in Figure 2 along the $z$-axis 
provides a natural 
source of CRs just where the 
cool radial filaments form.  
Although we do not compute the passively evolving magnetic field, 
it is expected to be stronger in denser thermal filaments 
and longitudinally oriented along the cool filaments 
due to the radial expansion of the filament flow along the $z$-axis. 

The 2D cylindrical nature of our solution requires that all 
cavities and CR activity occur along the $z$-axis.
The relatively small number of double-double radio sources 
(Schoenmakers et al. 2000) indicates that successive 
CR jets can move along the same direction, 
as assumed here.
Remarkably, the X-ray thermal filament in Virgo 
(Forman et al. 2007)
that formed by a cavity $\sim 10^8$ years ago 
(Mathews \& Brighenti 2008a) 
and its associated radio lobe 
lie in a different direction than the younger non-thermal 
jet in the central galaxy M87.
Another indication that CR energy flows out 
in many different directions is the azimuthal variation 
of successive generations of 
X-ray cavities in Perseus and other clusters. 

It is expected that the mass outflow 
due to CR buoyancy in Figure 1 
also transports iron  
produced by Type Ia supernova inside the central galaxy 
into a quasi-spherical region 50-100 kpc in radius 
as observed in 
many groups and clusters (De Grandi et al. 2004).
The outflows described here can also carry 
short-lived dust out to 5-10 kpc into the hot gas 
where its far-infrared emission is observed in many 
nearby galaxy groups 
(Temi, Brighenti, \& Mathews 2007).
Radial entropy variations qualitatively similar to those in Figure 1 
have been observed in galaxy groups 
(Gastaldello et al. 2007; Sun et al. 2008).
Finally, the outward mass transport can create an inflection in 
the density slope as in Figure 1, resembling the double-structured 
character of many density profiles observed.

\vskip0.1in
Studies of the evolution of hot cluster gas
at UC Santa Cruz are supported by
NASA and NSF grants for which we are very grateful.

\clearpage
\begin{figure}
\centering
\vskip2.7in
\hspace{2.in}
\includegraphics[bb=262 275 550 707,scale=0.8,angle= 270]
{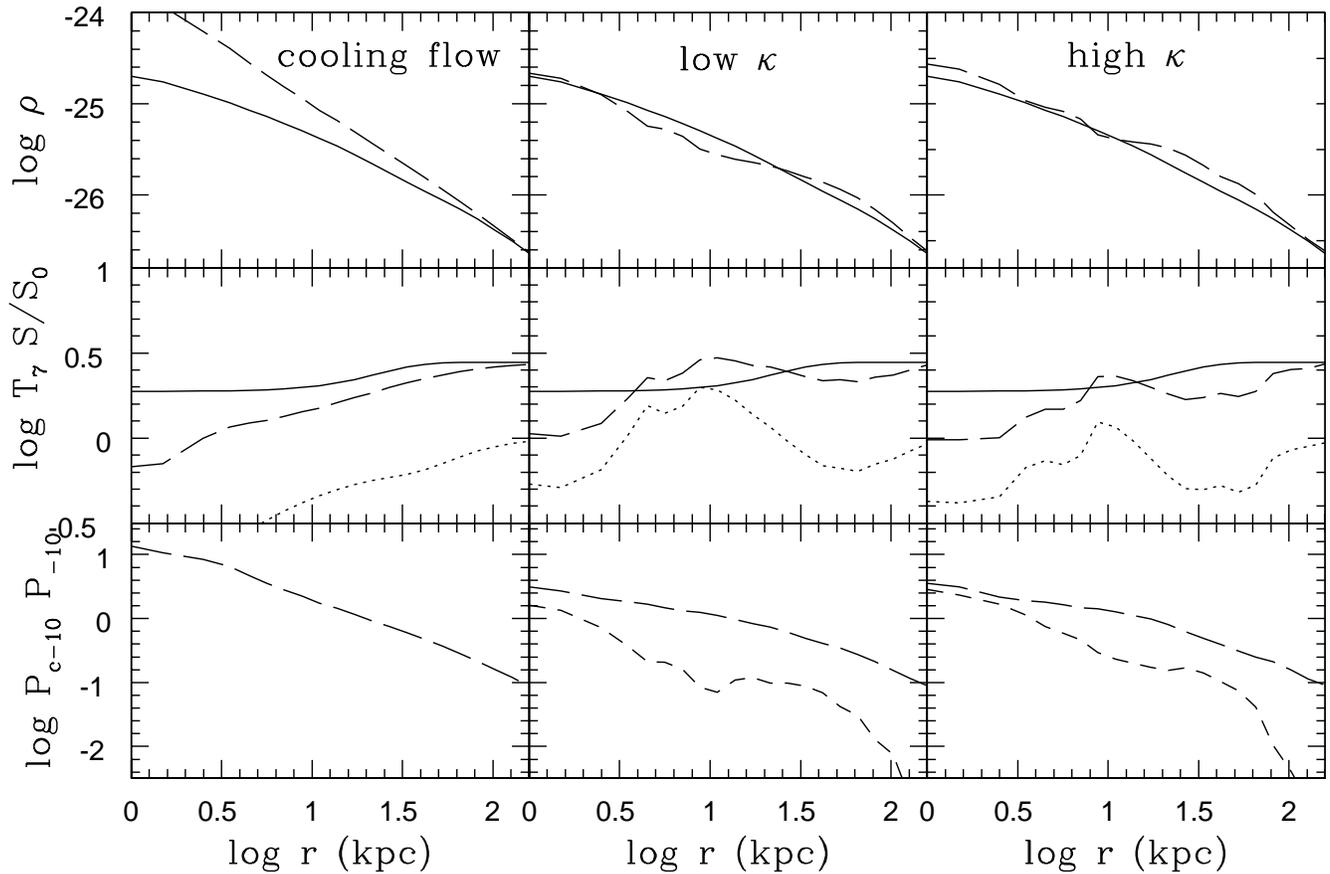}
\caption{
Radial distribution of gas and CRs in 
Virgo after 3 Gyrs of intermittent CR 
injection. 
Columns from left to right show 
a cooling flow without CRs, 
the same flow with weakly then strongly diffusing 
CRs. 
{\it Top row:} initial (observed) gas density 
({\it solid lines})
and density after 3 Gyrs 
({\it long dashed lines}).
{\it Central row:} initial (observed) gas temperature 
({\it solid lines}),
and after 3 Gyrs
({\it long dashed lines}) (in units of $10^7$ K) and 
the change $\log S/S_0$ of the gas entropy after 3 Gyrs
({\it dotted lines}).
{\it Bottom row:} pressure of gas  
({\it long dashed lines}) 
and CRs 
({\it short dashed lines}) after 3 Gyrs 
(units of $10^{-10}$ dynes cm$^{-2}$).
}
\label{f1}
\end{figure}

\clearpage
\begin{figure}
\vskip.5in
\centering
\hspace{2.in}
\includegraphics[bb=202   5 490 437,scale=1.0,angle= 0]
{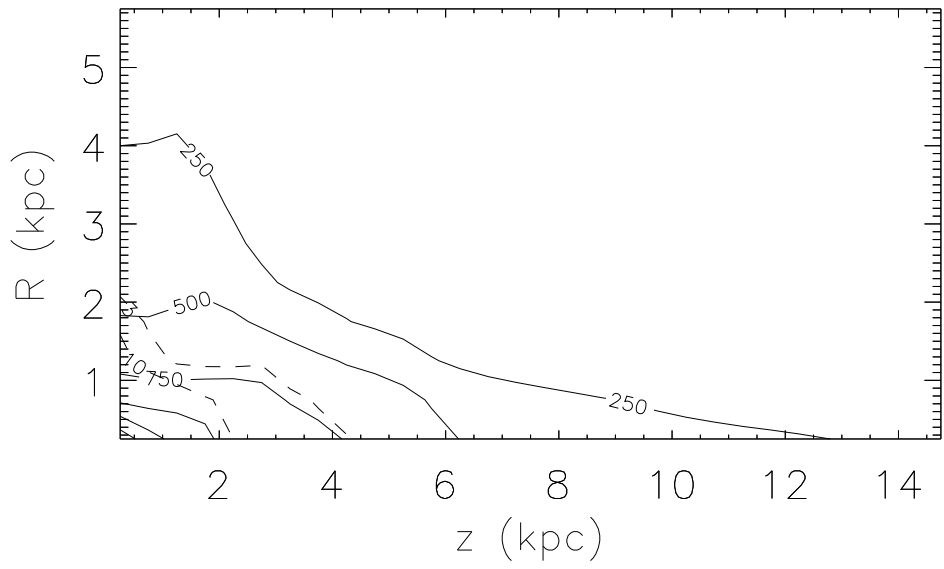}
\vskip.5in
\caption{
Distributions after 3 Gyrs of CR energy density 
$e_c(R,z)$ in units of $10^{-12}$ erg cm$^{-3}$ 
({\it solid lines}) and cooled mass in 
$10^8$ $M_{\odot}$
({\it dashed lines}).
}
\label{f2}
\end{figure}

\end{document}